\documentclass[12pt]{article} 
\usepackage{amsfonts}
\usepackage{amsmath, graphicx, epsfig,psfrag, verbatim}
\usepackage{eucal,  amssymb,amsmath,amsthm,graphicx, amsfonts, latexsym,xcolor}
\usepackage{caption}
\usepackage{subcaption}
\usepackage[utf8]{inputenc}

\begin{document} \parskip=5pt plus1pt minus1pt \parindent=0pt
\title{Improving the use of social contact studies in epidemic modelling}
\author
{Tom Britton$^{1\ast}$ and Frank Ball$^{2}$ \\
\\
\normalsize{$^{1}$Department of Mathematics, Stockholm University,  Sweden}\\
\normalsize{$^{2}$School of Mathematical Sciences, University of Nottingham,  UK.}\\
}
\date{\today}
\maketitle

\begin{abstract}
Social contact studies, investigating social contact patterns in a population sample, have been an important contribution for epidemic models to better fit real life epidemics. A contact matrix $M$, having the \emph{mean} number of contacts between individuals of different age groups as its elements, is estimated and used in combination with a multitype epidemic model to produce better data fitting and also giving more appropriate expressions for $R_0$ and other model outcomes. However, $M$ does not capture \emph{variation} in contacts \emph{within} each age group, which is often large in empirical settings. Here such variation within age groups is included in a simple way by dividing each age group into two halves: the socially active and the socially less active. The extended contact matrix, and its associated epidemic model, empirically show that acknowledging variation in social activity within age groups has a substantial impact on modelling outcomes such as $R_0$ and the final fraction $\tau$ getting infected. In fact, the variation in social activity within age groups is often more important for data fitting than the division into different age groups itself. However, a difficulty with heterogeneity in social activity is that social contact studies typically lack information on if mixing with respect to social activity is assortative or not, i.e.\ do socially active tend to mix more with other socially active or more with socially less active? The analyses show that accounting for heterogeneity in social activity improves the analyses irrespective of if such mixing is assortative or not, but the different assumptions gives rather different output.
Future social contact studies should hence also try to infer the degree of assortativity of contacts with respect to social activity.

\end{abstract}

\footnotetext[1]{Stockholm University, Department of Mathematics, Sweden. E-mail: tom.britton@math.su.se}


\section*{Introduction}\label{sec-Intro}
Epidemic models have a long history starting with assuming a completely homogeneous community, followed by many steps towards more realism: introducing stochasticity, incorporating household structure, dividing the population into different types of individual, modelling the effect of preventive measures and more, e.g.\ \cite{DHB13}. Such modelling extensions only contribute to the analysis of real world epidemics if fitted to data. One such important contribution has been the use of social contact studies in order to incorporate contact heterogeneities into the epidemic model. This began 15 years ago with the important POLYMOD study \cite{M08} in 8 European countries, and has continued with many social contact studies in different countries and under different settings, see \cite{SCD} where many such data sources are stored. During the Covid-19 pandemic, social contact studies were performed during several stages in the pandemic (including the CoMix project, see \cite{SCD}), thus also giving information on how contact patterns changed over the pandemic as an effect of restrictions, voluntary changes and vaccination (e.g. \cite{CWG20}, \cite{L21}).

Contact studies are typically based on random samples of individuals for which individual covariates, such as age, gender, household information are collected. The individuals are then asked to record each close contact (defined e.g.\ by being within arm-length to an individual for at least one minute) during a 24 hour period, and also record their age, whether a household member or not, and similar. Such a data set of e.g.\ 1000-2000 individuals hence consists of one row for each contact of each individual and several columns with information about the contact. The most common way to improve the epidemic model and its statistical analysis from contact studies has been to extract what is known as the age-dependent mean contact matrix $M=[\alpha_{ij}]$. For this the population is first divided into different age groups, and element $\alpha_{ij}$ is then defined as the \emph{mean} number of contacts that an individual in age group $i$ has with individuals of age group $j$ (during one day).
Nearly all such contact studies show two important features: 1) that younger age groups tend to have more contacts in total than older age groups, and 2) that all age groups tend to have most contacts with individuals of the same (or at least similar) age group, the latter is referred to assortative mixing with respect to age \cite{HCM19}.

The matrix $M$ is then incorporated into a multitype epidemic model, rather than assuming homogeneous mixing (or some more arbitrary mixing assumption between age groups) and it it has been shown to give a better statistical fit to incidence data (e.g.\ \cite{WTK06}).

The separation of the community into different age groups
 has the effect of making individuals of the same age group resemble each other more with respect to contact pattern. However, there still remains heterogeneity also within each age group: not everyone in age group $i$ has $\alpha_{ij}$ contacts with individuals of age group $j$. The more heterogeneity that remains within the age groups the cruder is the approximation of the multitype epidemic which assumes that all individuals of the same age group mix in a similar way.

In the present paper we hence define a new multitype epidemic model which tries to capture also this remaining heterogeneity, after separating individuals into different age groups. We do this by dividing each age group in the original age-group division into two halves: those with low socially activity and the other half being those with a higher number of social contacts. Each age group $i$ is hence divided into two groups, $(i,L)$ and $(i,H)$, for Low and High individuals in age group $i$. For this new group classification we know how many contacts on average an $(i,L)$-individual has with $j$-individuals (and similarly how many $(i,H)$-individuals have with $j$-individuals) but we do not know which fraction of those contacts are with $(j,L)$-individuals and which fraction are with $(j,H)$-individuals. For this reason we consider the two extreme situations: the fully assortative case where $(i,H)$-individuals prioritize $(j,H)$-individuals (and $(i,L)$-individuals prioritize $(j,L)$-individuals), and the fully disassortative case where $(i,H)$-individuals prioritize $(j,L)$-individuals as much as possible (contacts are symmetric so the number of contacts by $(i,H)$-individuals to $(j,L)$-individuals must equal the number of contacts by $(j,L)$-individuals to $(i,H)$-individuals). Beside the two extreme situations of fully assortative and fully disassortative, we consider an intermediate situation known as proportionate mixing (with respect to social activity) in which contacts are selected in the same way for socially active and socially less active individuals.

We also consider a simpler epidemic model in which we neglect age completely and instead only divide the population according to how many contacts individuals have in total: for example dividing the community into two halves, those having a high number of social contacts and those having low social activity. For this stratification into different types of individual reflecting social activity only, we derive a matrix of mean number of contacts between different groups of individuals. However, also in this simpler model, the data lack information on how the total number of contacts of an individual divide into its contacts with high-active and low-active individuals. Also here we consider the two extreme situations where socially active individuals \emph{only} have contact with socially active individuals (assortative mixing with respect to social activity) and the opposite situation where socially active individuals tend to prioritize contacts with low-social-activity individuals (denoted disassortative mixing with respect to social activity), and the intermediate proportionate mixing case. Hence again we end up with three models to analyse.

\section*{Data and epidemic models}
As described above we make use of empirical data sets from contact surveys. More precisely we have chosen to analyse three contacts surveys for which the data is publicly available at http://www.socialcontactdata.org/data/. We choose three data sets representing contacts in Belgium during 2010 \cite{W12}, in France during 2012 \cite{B11} and in Vietnam during 2007 \cite{H11}, the latter being chosen to see differences between contacts in Western societies and low-income countries. All data sets were intentionally chosen from before the Covid-19 pandemic, since studies from the pandemic often focused on changes over time as an effect of preventions, something which is not considered here.

For each of the three studied contact data sets, we analyse properties of the Homogeneous epidemic model with no heterogeneous structure ($Hom$), the multitype epidemic model with heterogeneity with respect to age only ($A$), the multitype epidemic model with heterogeneity with respect to social activity only ($S$),  and the multitype epidemic model with heterogeneity with respect to both age and social activity within age groups ($AS$). For the models with heterogeneity with respect to social activity ($S$ and $AS$), we consider both the fully assortative case (socially active only have contact with other socially active), the intermediate case referred to as proportionate mixing and the fully disassortative case (socially active mainly have contact with socially less active). The ages are divided into 7 different age groups: 0-5, 6-12, 13-18, 19-24, 25-44, 45-64 and 65+. Social activity on the other hand, is divided into only two classes: the 50\% least socially active and the 50\% most socially active individuals, in the entire population for the S-model and within each age group for the AS-model. For the assortative models we in fact have a tiny fraction $\epsilon$ of mixing between the high and low social activity groups in order not to make the two groups completely isolated from each other (see Supp.~Mat., Section 4.1).

For each of the three social contact studies and each epidemic model, we consider the corresponding stochastic multitype epidemic model (see methods and e.g. \cite{AB00}, Chapter 6) and consider a population size tending to infinity. For each model we compute the basic reproduction number $R_0$, the limiting probability of a major outbreak $\rho$, and the limiting final fraction getting infected in case of a major outbreak $\tau$, as a function of the overall transmissibility (measured by the transmission probability $p$).

\section*{Results}

\begin{figure}
\centering
\begin{subfigure}[t]{0.49\textwidth}
\centering
\includegraphics[width=1\linewidth]{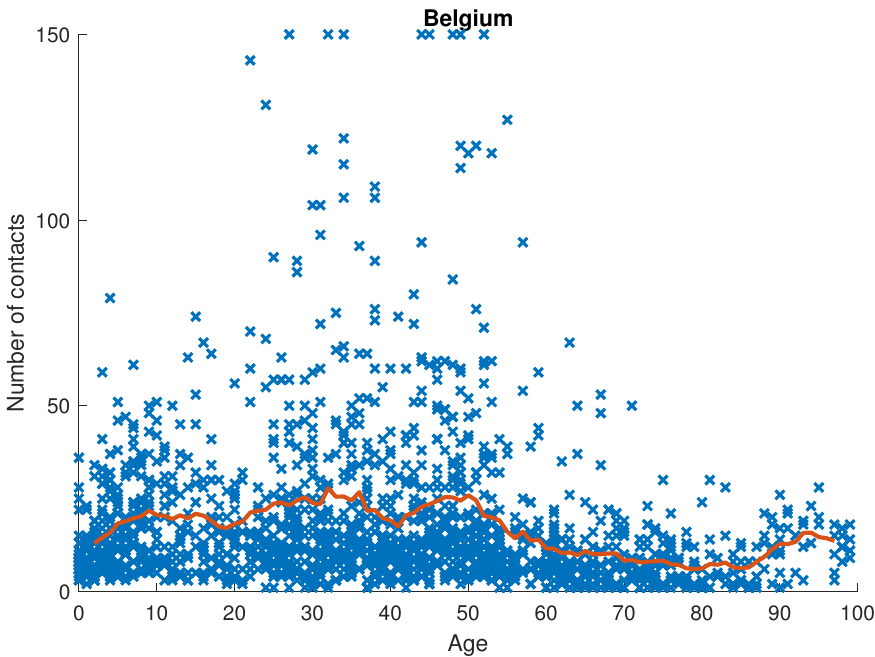}
\end{subfigure}
\begin{subfigure}[t]{0.49\textwidth}
\centering
\includegraphics[width=1\linewidth]{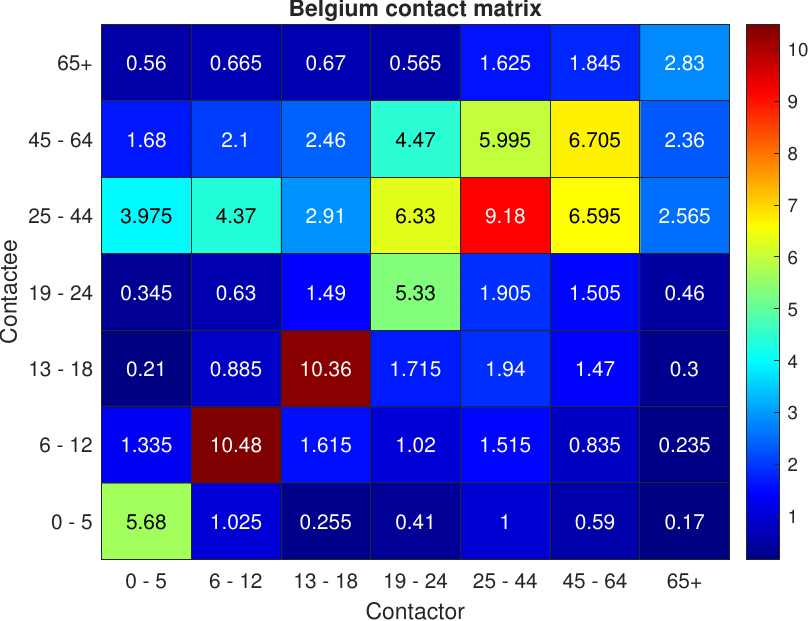}
\end{subfigure}
\caption{Belgian data  \cite{W12}. Scatter plot of the total number of contacts of individuals as a function of age and 4-year moving average as red curve; 8 observations with larger values are truncated and set to 150 (left panel).  Heatmap representing the contact matrix $M$ between different age groups (right panel).}
\label{fig_belgium_data}
\end{figure}

We start by analysing the Belgian social contact study \cite{W12} consisting of 1744 individuals.  In Figure \ref{fig_belgium_data} we plot the total number of contacts (8 observations are truncated at 150 for better visibility) for all individuals as a function of age, and also a heatmap of the contact matrix $M$ where individuals are grouped into 7 different age-cohorts.

From the scatter plot it is seen that there is big variation in the number of contacts, but also that there is an age effect with highest overall number of contacts in the age group 20-30 year old. In the heatmap matrix it is seen which age groups individuals have most contacts with, and there is quite strong assortativity (with respect to age) in that individuals mix the most with individuals of similar age (please note that the age-ranges varies between groups somewhat blurring the diagonal (assortative) mixing proporty).

We next analyse several different epidemic models using the Belgian data, fixing all parameters except the overall transmissibility $p$, and compute $R_0$ and the final outcome size $\tau$ of a major outbreak under different model assumptions concerning social contact.

\begin{figure}
\centering
\begin{subfigure}[t]{0.32\textwidth}
\centering
\includegraphics[width=1\linewidth]{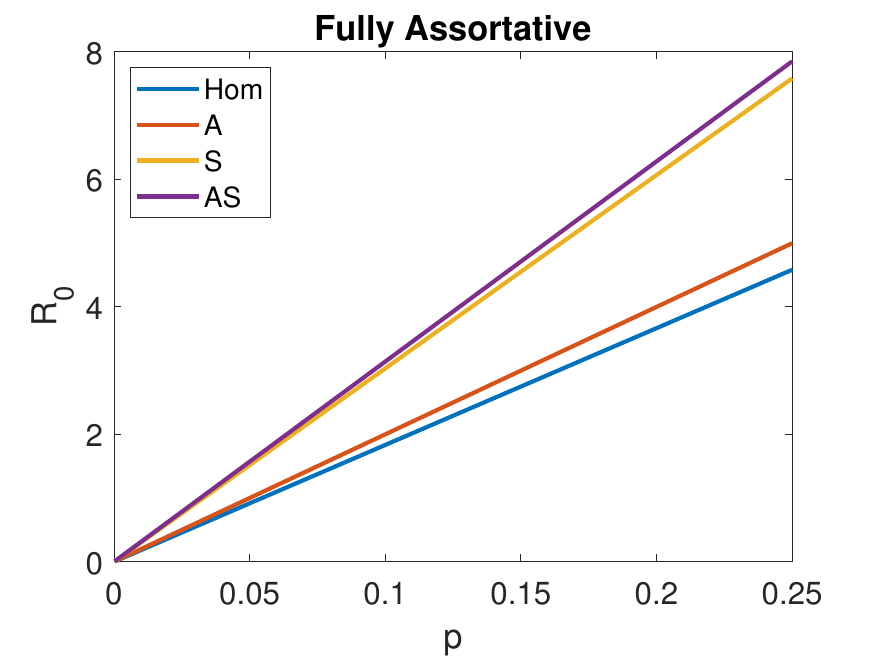}
\end{subfigure}
\begin{subfigure}[t]{0.32\textwidth}
\centering
\includegraphics[width=1\linewidth]{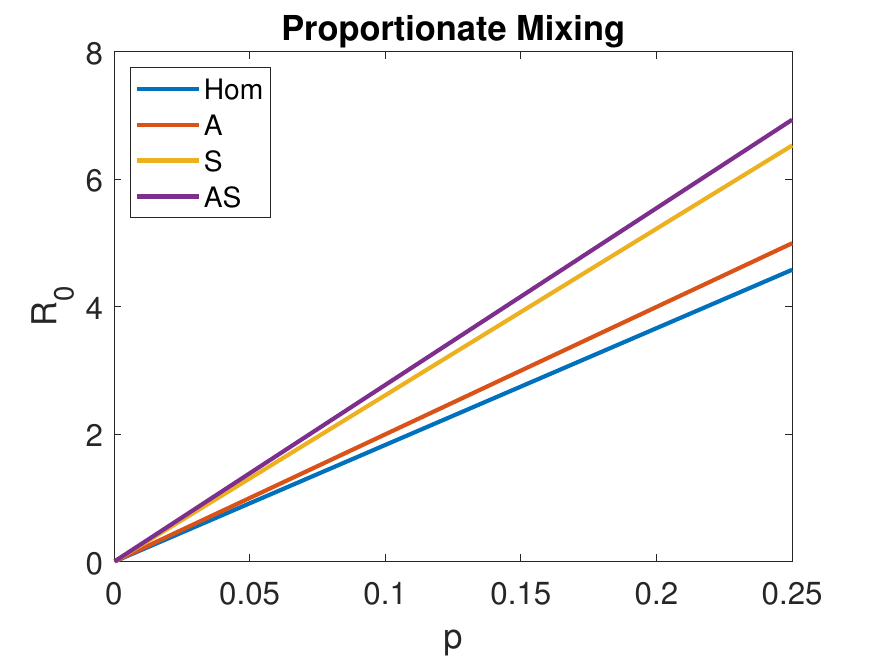}
\end{subfigure}
\begin{subfigure}[t]{0.32\textwidth}
\centering
\includegraphics[width=1\linewidth]{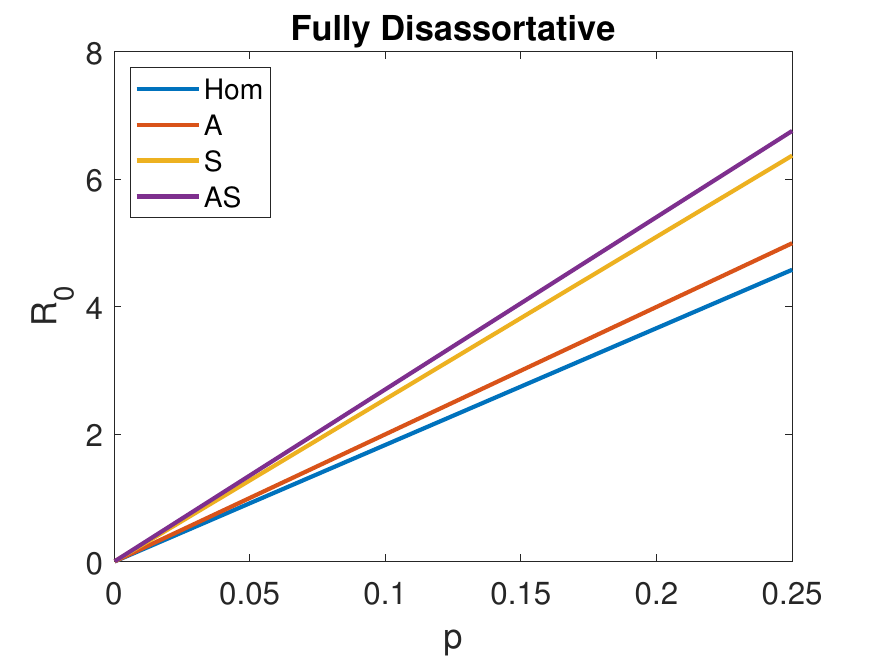}
 \end{subfigure}
 \caption{Plot of the basic reproduction number $R_0$, as a function of the per contact transmission probability $p$, for the Belgian contact study  \cite{W12}, analysed using epidemic models acknowledging no heterogeneity ($Hom$), heterogeneity with respect to age only ($A$), heterogeneity with respect to social activity only ($S$), and heterogeneity with respect to both age and social activity ($AS$), and considering mixing with respect to social activity to be fully assortative (left panel), proportionate mixing (center panel) and fully disassortative right (panel).}
\label{fig_belgium_R}
\end{figure}

In Figure \ref{fig_belgium_R} we plot $R_0$ for the different models: $Hom$, $A$, $S$ and $AS$. The right panel shows $R_0$ assuming complete assortativity with respect to mixing between social activity groups, the middle panel assumes proportionate mixing with respect to social activity, and the right panel assumes maximal disassortativity with respect to mixing between social activity groups. Note that the $Hom$- and the $A$-models do not contain social activity groups and hence only have one version, as opposed to the $S$- and $AS$-models which have three versions: assortative, proportionate mixing and disassortative. It is seen in the figure that for all models $R_0$ is linear in the overall transmissibility. In each panel, (i) the $AS$-model gives the highest $R_0$ and the $Hom$-model gives lowest $R_0$ and (ii) the $S$-model is closest to the $AS$-model and the $A$-model is closest to the $Hom$-model. Further, it is seen that assuming assortative mixing with respect to social activity (left panel) gives the highest $R_0$ and the disassortative assumption (right panel) gives the lowest $R_0$.

\begin{figure}[h]
\centering
\begin{subfigure}[t]{0.32\textwidth}
\centering
\includegraphics[width=1\linewidth]{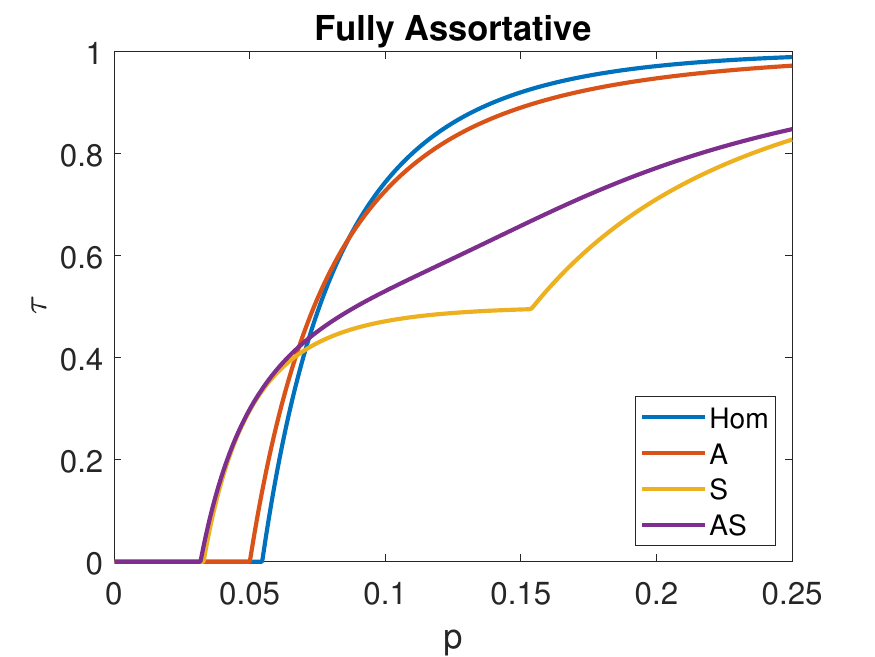}
\end{subfigure}
\begin{subfigure}[t]{0.32\textwidth}
\centering
\includegraphics[width=1\linewidth]{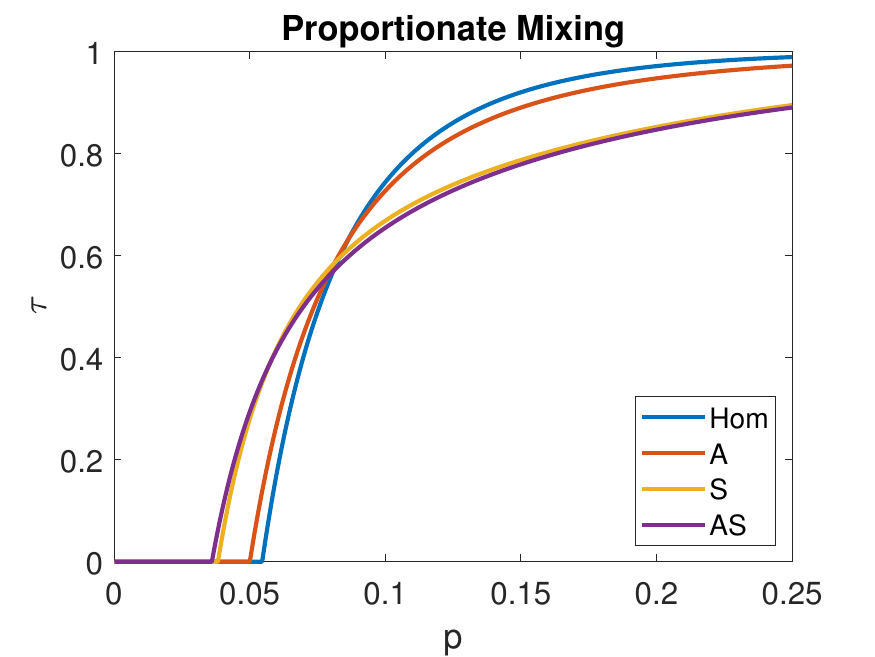}
\end{subfigure}
\begin{subfigure}[t]{0.32\textwidth}
\centering
\includegraphics[width=1\linewidth]{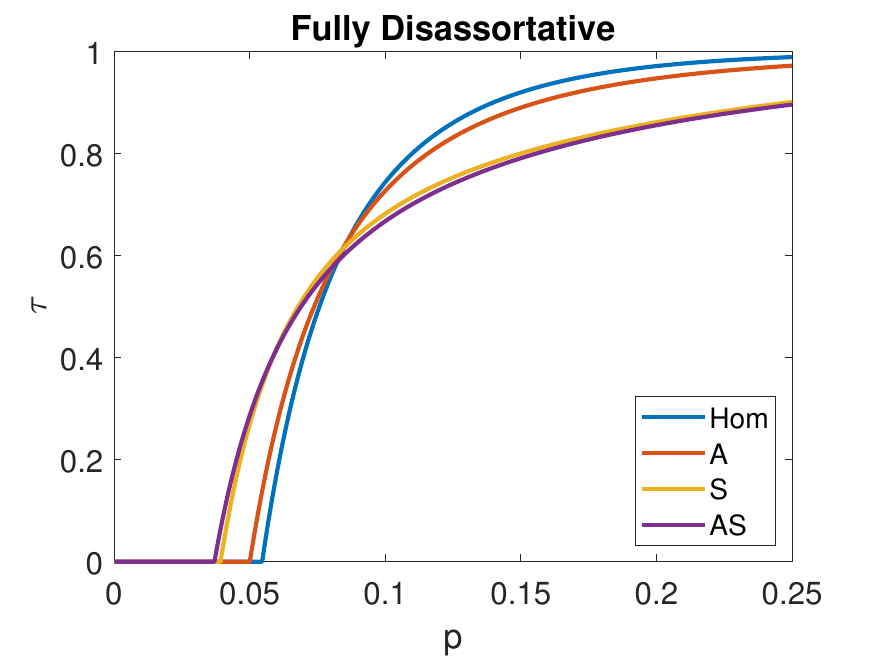}
 \end{subfigure}
 \caption{Plot of the final size $\tau$, as a function of the per contact transmission probability $p$, for the Belgian contact study  \cite{W12}, analysed using epidemic models acknowledging no heterogeneity (Hom), heterogeneity with respect to age only (A), heterogeneity with respect to social activity only (S), and heterogeneity with respect to both age and social activity (AS), and considering mixing with respect to social activity to be fully assortative (left panel), proportionate mixing (center panel) and fully disassortative right (panel).}
\label{fig_belgium_tau}
\end{figure}

Next, in Figure \ref{fig_belgium_tau} we plot the final size $\tau$ for the different models: $Hom$, $A$, $S$ and $AS$. Similar to above, the right panel assumes complete assortativity, the middle panel proportionate mixing, and the right panel disassortativity (with respect to social activity). It is seen in each panel, that the $A$-model (taking age into account) resembles the $Hom$-model (assuming no heterogeneities), whereas the $S$-model (acknowledging heterogeneity with regards to social activity but not age) is fairly similar to the full $AS$-model allowing heterogeneities both with respect to age and social activity. None of the four models is of course ``true'', but it is reasonable to believe that the $AS$-model lies closest to reality in that it allows for heterogeneities with respect to both age and social activity. As a consequence, including heterogeneity owing to social activity ($S$) is more important than acknowledging heterogeneity with respect to age ($A$).

Another observation is that, in each panel, the $Hom$- and $A$-models give smaller outbreaks (compared with the $S$ and $AS$-models) for low overall transmissibility $p$ whereas the opposite holds true when $p$ is large. An intuitive explanation to this is that strong heterogeneity ($S$ and $AS$) ``helps'' an epidemic to take off when $p$ is small, but that strong heterogeneity also ``helps'' some individuals to escape infection when $p$ is large.

If instead we compare the different panels in Figure \ref{fig_belgium_tau} we note that the final sizes for the $S$- and $AS$-models are different in the three panels (the $Hom$- and $A$-models are identical in the three panels since they do not include social activity considerations). In particular the left assortative panel is different from the others: the final size is smaller when $p$ is large, and the $S$-model has a type of ``bump'' at around $p=0.015$ which is an artefact from having two social activity groups. It is not obvious which of the three panels, assortative, proportionate mixing or disassortative, is the most reasonable but empirical studies, e.g.\ \cite{ZLH20}, tend to indicate that mixing with respect to social activity is assortative, albeit not completely, suggesting that somewhere between left and centre panel is closest to reality. However, for all three panels, and hence most likely also for intermediate situations, the $S$-model is much closer to the full (=best) $AS$-model. Similar conclusions to those from Figure \ref{fig_belgium_tau} for the final size $\tau$ apply also for the major outbreak probability $\rho$ (see Supp.~Mat.). As a numerical example we assume $p=0.1$ and full assortativity. Then most modellers would have used the $A$-model and conclude that $R_0= 2.00 $ and that the epidemic would result in an outbreak infecting the fraction $\tau=72.71\%$, whereas if the better $AS$ model would have been used the conclusion would instead have been $R_0=3.14$ and $\tau=53.06\%$ (so a higher $R_0$ but smaller final size). The $S$-model comes much closer to the $AS$-model: $R_0=3.03$ and $\tau=47.13\%$.

Note that the $S$-model lies closer to the $AS$-model than the $A$-model even though we have 7 age-groups and only two social activity groups, thus allowing for more heterogeneity for age. In Sup.~Mat.\ we show similar plots when dividing also social activity into 7 different levels, and then differences between $Hom$ and $A$ on the one side and $S$ and $AS$ on the other are even more pronounced.

\begin{figure}
\centering
\begin{subfigure}[t]{0.32\textwidth}
\centering
\includegraphics[width=1\linewidth]{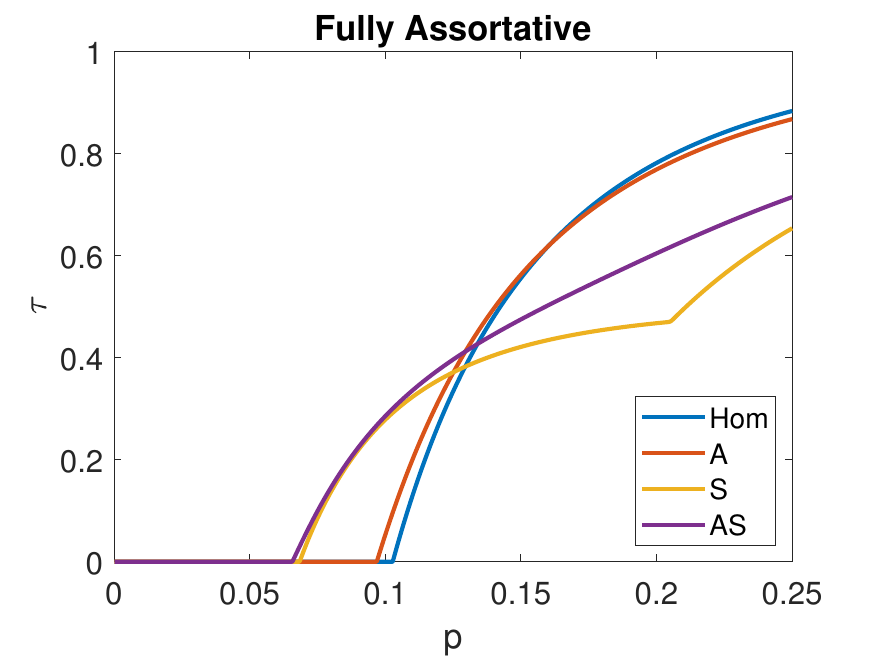}
\end{subfigure}
\begin{subfigure}[t]{0.32\textwidth}
\centering
\includegraphics[width=1\linewidth]{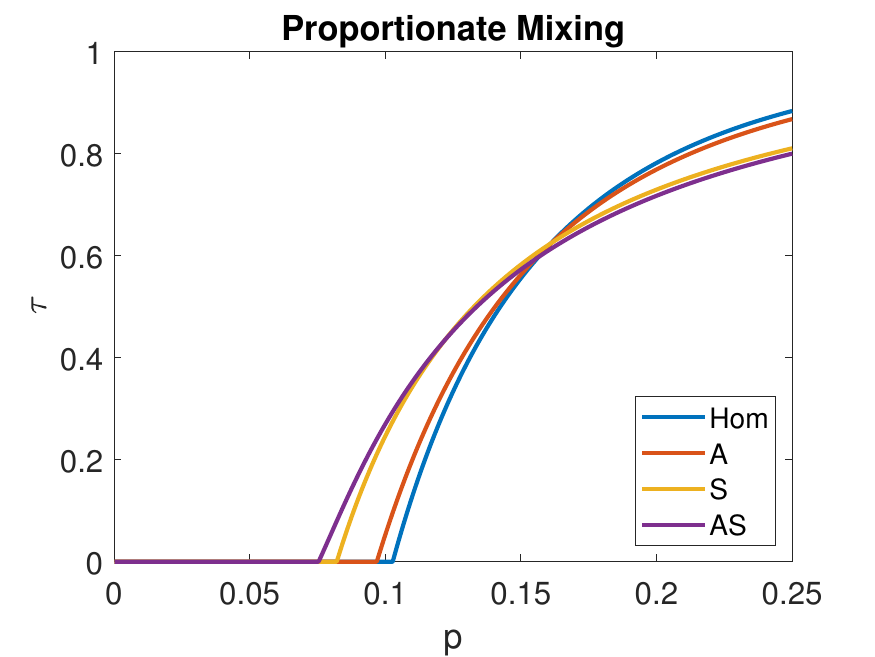}
\end{subfigure}
\begin{subfigure}[t]{0.32\textwidth}
\centering
\includegraphics[width=1\linewidth]{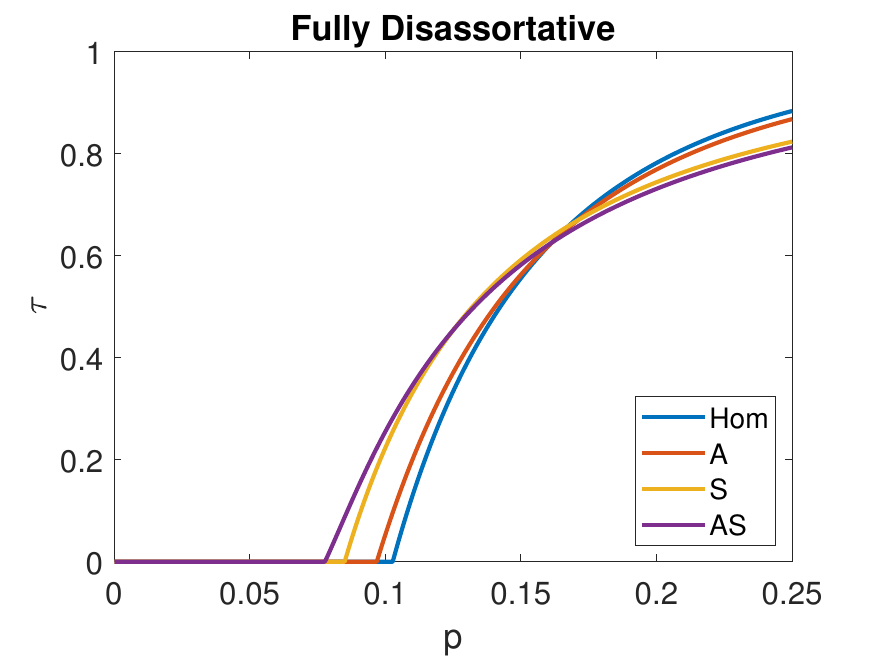}
 \end{subfigure}
 \caption{Plot of the final size $\tau$, as a function of the per contact transmission probability $p$, for the French contact study  \cite{B11}. Same panels and assumptions as in Figure \ref{fig_belgium_tau}.}
\label{fig_french_tau}
\end{figure}

We have also done similar comparative analyses for two other social contact studies, from France \cite{B11} and  Vietnam  \cite{H11}, which are described more briefly. The conclusions from the French social contact study are very similar to those from the Belgian analysis, as can be seen in Figure \ref{fig_french_tau} where we plot the final fraction $\tau$ getting infected as a function of the overall transmissibility $p$ for the different model assumptions $Hom$, $A$, $S$ and $AS$, under different assumptions on assortativity with respect to social activity.

\begin{figure}
\centering
\begin{subfigure}[t]{0.49\textwidth}
\centering
\includegraphics[width=1\linewidth]{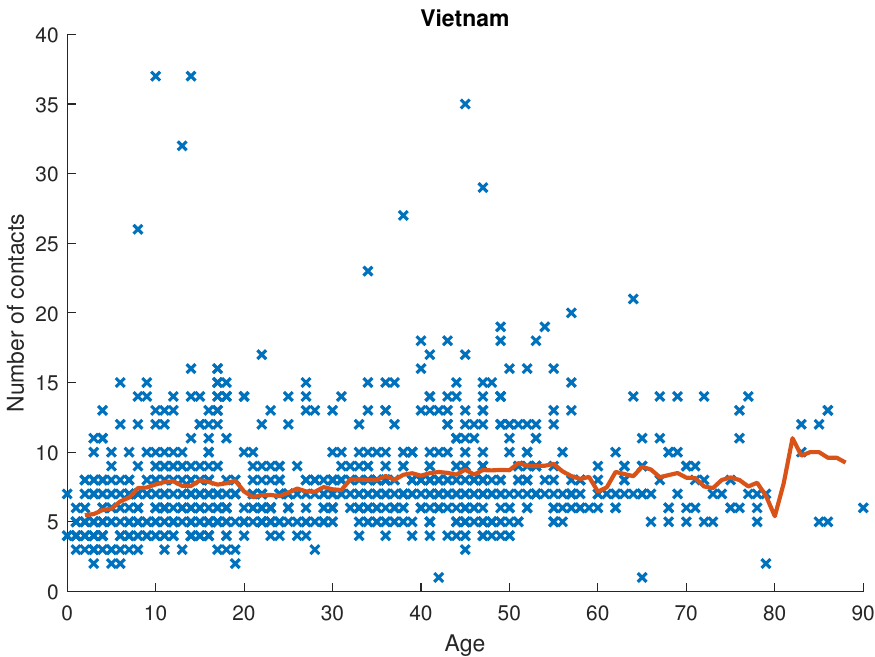}
\end{subfigure}
\begin{subfigure}[t]{0.49\textwidth}
\centering
\includegraphics[width=1\linewidth]{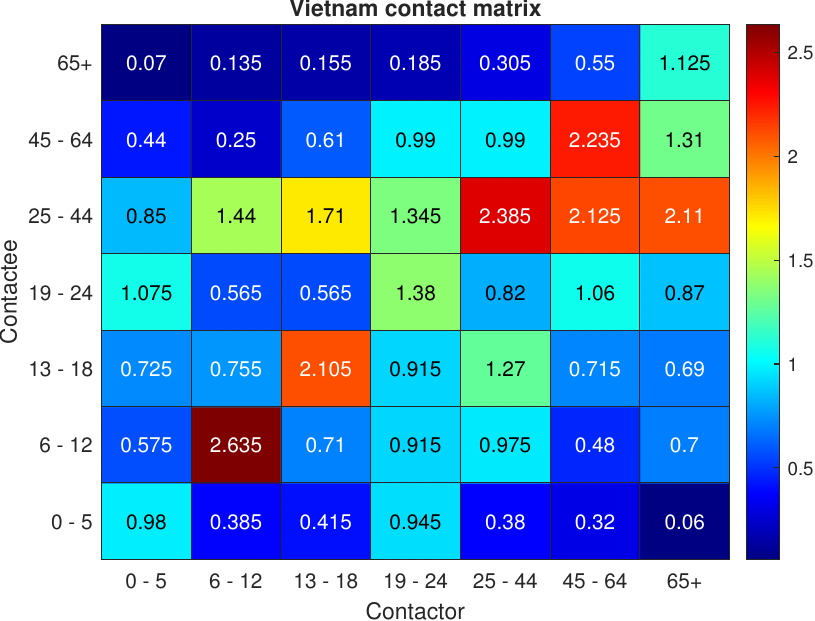}
\end{subfigure}
\caption{Vietnamese data   \cite{H11}. Scatter plot of the total number of contacts of individuals as a function of age, red curve showing the moving average (left panel) and heatmap representing the contact matrix $M$ between different age groups (right panel).}
\label{fig_vietnam_data}
\end{figure}

The Vietnamese data on the other hand show some differences. The main difference of the Vietnamese social contact study is that the number of contacts varies much less, both within and between age groups, compared to the Belgian and French data. In Figure \ref{fig_vietnam_data} a scatter plot of the number of contacts is shown as well as a heatmap of the contact matrix with respect to age. It is seen that the variation in number of contacts is \emph{much} smaller than in the Belgian data (Figure \ref{fig_belgium_data}), and that there is still some assortativity for mixing with respect to age with high values on the diagonal.

\begin{figure}
\centering
\begin{subfigure}[t]{0.32\textwidth}
\centering
\includegraphics[width=1\linewidth]{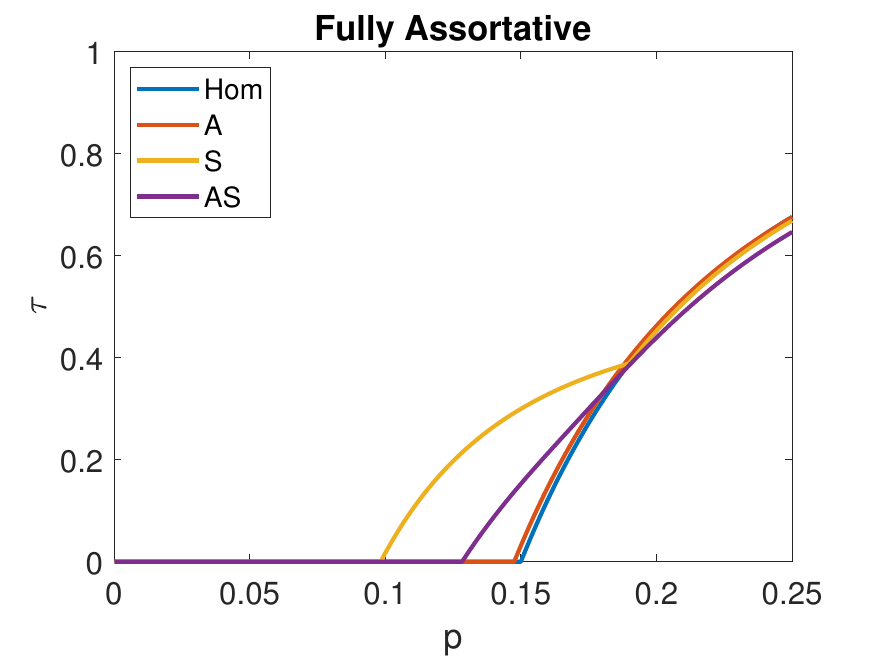}
\end{subfigure}
\begin{subfigure}[t]{0.32\textwidth}
\centering
\includegraphics[width=1\linewidth]{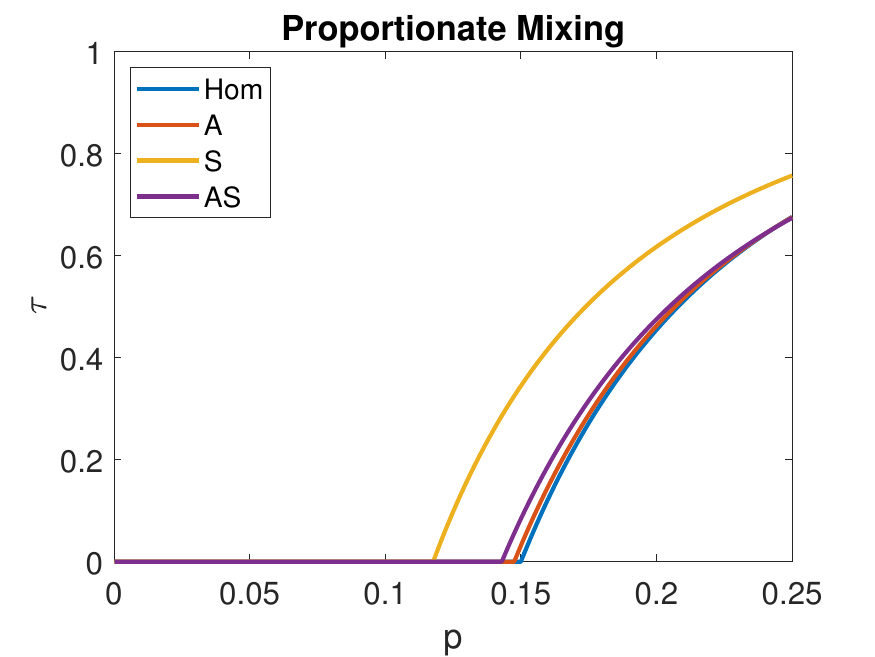}
\end{subfigure}
\begin{subfigure}[t]{0.32\textwidth}
\centering
\includegraphics[width=1\linewidth]{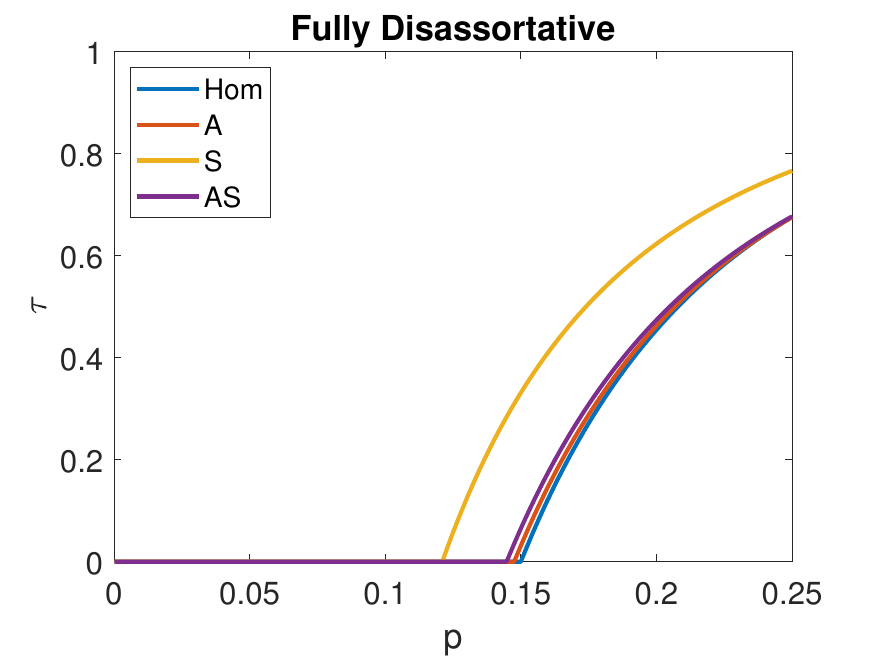}
 \end{subfigure}
 \caption{Plot of the final size $\tau$, as a function of the per contact transmission probability $p$, for the Vietnamese contact study  \cite{H11} . Same panels and assumptions as in Figure \ref{fig_belgium_tau}.}
\label{fig_vietnamese_tau}
\end{figure}

When comparing the potential of outbreaks for different models the pattern is different from the Belgian and French social contact studies, see Figure \ref{fig_vietnamese_tau}. The only model that sticks out now is the $S$-model. Since there is very little variation between age groups and also within age groups the homogeneous model ($Hom$) might suffice. The reason why the $S$-model sticks out is probably because it has only two groups of individuals, and even if the variation in number of contacts, the mean in the high and low social activity groups differ to some extent (mean about 10 compared to mean of about 5).

\section*{Discussion}

The two main insights from our analyses are that 1) heterogeneity in social activity is more important than that owing to age when determining what will happen if an epidemic outbreak takes place in a community, and 2) whether or not mixing is assortative with respect to social activity is also an important factor for determining epidemic features such as $R_0$ and the final size $\tau$, but current social contact studies lack information on the degree of such assortativity. The first insight implies that epidemic models using social contact studies should include also heterogeneity in social activity within age groups in the epidemic model to improve analyses. The second insight calls for attention in future social contact studies to investigate also to what degree social mixing is assortative with respect to social activity.

The conclusions are based on three social contact studies from Belgium, France and Vietnam, and the significance of modelling heterogeneity in social activity within age groups was smallest in the Vietnamese study where variation in number of contacts was the smallest. To allow for heterogeneity within age groups is most important when variation in contacts is large also within age groups. But the paradigm in social mixing seems to be that social activity is highly heterogeneous (super-spreaders, scale-free social networks, ...), so we believe it is rather the rule than the exception that social activity is highly variable both within and between age groups in communities.

Earlier modelling using social contact studies made use of the contact matrix $M$ having the \emph{mean} number of contacts between different age-cohorts as its element. But even if there were no systematic differences between individuals of the same age group a social contact study would of course not have all individual contacts of a certain age group being identical -- there would clearly be some variation also then. A relevant question is therefore if the observed variation within age groups is systematic or more noise-related. We have two strong empirical indicators that variation in contacts \emph{is} systematic. The first is that the number of contacts is strongly over-dispersed also within age groups. The variance of the number of contacts divided by the mean number of contacts within each age group vary between 11.4 and 54.3 for the Belgian data and between 13.6 and 26.6 for the French data (but only between 3.7 and 6.8 for the Vietnamese data). If there were no systematic differences between individuals a Poisson distribution seems plausible and then the ratio would lie around 1. The second strong indicator that there are systematic differences in number of contacts also within age groups comes from the French data where 278 individuals in the study were measured on four different days.  (The data comprise two waves, with each wave consisting of two successive days.)  By analysing these 278 individuals it was found that the correlation between the numbers of contacts on the different days was estimated to lie in the range $r=0.30-0.52$, a strong positive correlation, whereas if the variation was just noise one would expect no correlation. Hence, there are very strong indications of systematic variation in number of contacts within age groups, and this should be included in the modelling to improve analyses. It is however possible that the (systematic) heterogeneity within age groups is over-estimated in our analysis, since the separation of age groups into two halves, based on observed numbers of contacts, is affected by both systematic heterogeneities and noise. An important open question is hence how to disentangle these different sources of variation to improve the estimated systematic differences within age groups.

As mentioned earlier, social contact studies performed during the Covid-19 pandemic, including the CoMix initiative (\cite{SCD}), studied changes in social activity during the pandemic. Also there the focus was often on changes in the \emph{mean} number of contacts between different age groups, but it would be interesting to investigate if reduction in contacts come mainly from socially active individuals reducing their contacts (decreasing dispersion) or mainly from socially less active individuals reducing their contacts even further thus increasing dispersion, or a mixture of both. These scenarios would clearly impact the potential for disease spreading differently.

Our modelling focuses on the use of social contact studies and hence lacks several other relevant features in epidemic modelling, for example, seasonality, the effect of local structures such as households and workplaces, and immunity waning. Clearly such features are important for modelling conclusions to be more accurate.  However, we believe that our qualitative conclusions remain valid also in such more realistic models: when using social contact studies variation within age groups should preferably also be included in the modelling.

\small{

\section*{Methods}

In the Supp.~Mat.~we give an in-depth  description of how the contact survey data are used, and how they are incorporated into the multitype epidemic model and properties of  a multitype epidemic model -- here we give only a brief outline.

\subsection*{Contact survey data}

Each of the three social contact surveys (\cite{W12}, \cite{B11} and \cite{H11}) contains information about the sampled individuals and about the contacts they have during a given day. The data contains more information, but in the present analyses we only use the age of the sampled individuals and their contacts during the day in question: how many and the ages of each contact. For a given age group $i$, the contact matrix element $\alpha_{ij}$ is given by the mean number of contacts that $i$-individuals have with individuals in age group $j$.  This defines the contact matrix $M=(\alpha_{ij})$ used in the multitype epidemic model taking heterogeneity with respect to age into account ($A$-model).

For the epidemic model taking social activity into account but not age ($S$-model), we divide the sampled individuals into two halves: the 50\% fraction having most overall number of contacts and the remaining 50\% have fewer overall number of contacts. The mean $m_H$ among the group with High social activity is then computed and similarly $m_L$ is computed for the group having Low social  activity. As discussed earlier, the contact studies lack information about if the contacted individuals have High social activity or not, so this we have to hypothesize about. If we assume full assortativity with respect to social activity, then High individuals only have contact with other High individuals, and Low individuals only with other Low individuals. The contact matrix under this assumption hence has elements  $\alpha_{HH}=m_H$, $\alpha_{LL}=m_L$ and  $\alpha_{HL}=\alpha_{LH}=0$.

The fully disassortative model assumes that all contacts from Low individuals are with High individuals, but it is not possibly for all contacts of High individuals to be with Low individuals -- there are simply not enough contacts made by Low individuals, so the remaining contacts have to be within the High group. Consequently, the disassortative model has the following elements in its contact matrix: $\alpha_{HH}=m_H-m_L$, $\alpha_{HL}=\alpha_{LH}=m_L$ and $\alpha_{LL}=0$.

Finally, in the proportionate mixing assumption (of the $S$-model), each contact, irrespective of whether it comes from a High or Low individual, has the probability $p_H=m_H/(m_H+m_L)$ to be with a High individual, and the remaining probability $p_L=m_L/(m_H+m_L)$ to be with a Low individual. Hence, the contact matrix $M$ assuming proportionate mixing has elements $\alpha_{HH}= m_Hp_H$ , $\alpha_{HL}= m_Hp_L$, $\alpha_{LH}= m_Lp_H$ and $\alpha_{LL}= m_Hp_H$. We have thus defined the 2*2 contact matrices $M$ in the $S$ model under the three different mixing assumptions: assortative, disassortative and proportionate mixing. These are used in the multitype epidemic model described in the next subsection.

In the model taking heterogeneity with respect to both age and social activity into account ($AS$-model), we divide individuals of each age group $i$ into two: $(i,H)$ and $(i,L)$ being the 50\% with highest overall social activity and the 50\% with lowest overall social activity. It is known how many contacts $(i,H)$-individuals have with individuals in age group $j$ on average (and similarly for $(i,L)$-individuals), but just like in the $S$ model it is not known if these contacts are primarily with $(j,H)$- or $(j,L)$-individuals. Hence, also here we make three different assumptions: assortative mixing, proportionate mixing and disassortative mixing, all with respect to social activity. The details are given in Supp.\ Mat., Section 2, but each gives a contact matrix $M$ of dimension $(2*7)*(2*7)=14*14$ if there are 7 different age groups.

\subsection*{Multitype epidemic model}
In the previous subsection we outlined how the social contact study is used to produce a contact matrix $M$ under different assumptions. In the $S$-model $M$ is 2*2, in the $A$-model (explained earlier) the contact matrix has dimension 7*7 when there are 7 different age groups, and finally the $AS$-model has a contact matrix $M$ of dimension 14*14.

Whichever model is assumed we hence have some fixed number $k$ of types and it is known how many contacts with $j$-individuals of type $i$ have on average in a typical day: $\alpha_{ij}$. It is also assumed that we know the community fraction of the different types $\pi_1, \dots \pi_k$ and the overall community size $n$. In our application these are easy to find from National statistical bureaus: what is the country population and what fraction have age in given age cohorts.

A multitype SEIR epidemic model is defined for this setting as follows (see Supp.~Mat., Section 4, or \cite{AB00}, Chapter 6, for details). Individuals are at first susceptible, and someone who gets infected then has infectious contacts  randomly in time (possibly after a latent period) and after some (random) time in the infectious state they recover and become fully immune for the duration of the study period. During the infectious period $i$-individuals have on average $\alpha_{ij}$  contacts with $j$-individuals per day, where $M=[\alpha_{ij}]$ is the contact matrix described above and inferred from the social contact study. Let $\mu_I$ denote the mean duration of the infectious period and let $p$ denote the probability that an infectious contact results in infection, called the transmissibility of the disease in question. The mean number of infectious contacts (i.e.\ resulting in infection if the contacted person is susceptible) an $i$-individual has with $j$-individuals hence equals $\alpha_{ij} \mu_I p$. Without loss of generality we assume that $\mu_I=1$ (if not we can redefine the contact matrix $M$ in units of $\mu_I$ rather than per day, by multiplying all elements by $\mu_I$). This multitype epidemic model has been analysed extensively in the literature, see e.g.\ references in Supp.~Mat.~or \cite{AB00}, Chapter 6. It is known that the basic reproduction number $R_0$ is given by the largest eigenvalue of $pM$. Further, if the epidemic takes off (which may happen only if $R_0>1$) then the (random) final fractions getting infected of the different types converge to a deterministic limit as the population size $n$ tends to infinity. These limiting fractions $\tau_1,\dots, \tau_k$ are given as the unique strictly positive solution to the $k$ equations
$$
1-\tau_j = {\rm e}^{-p\sum_i^k \pi_i\tau_i \alpha_{ij}/\pi_j},\quad j=1,\dots , k.
$$

}

\section*{Acknowledgements}

We thank Sue Ball for assistance with processing the data. T.B.\ is grateful to the Swedish Research Council (grant 2020-0474) for financial support. Please contact Tom Britton for the Supplementary Material.


\begin{thebibliography}{00}


\bibitem{AB00} H Andersson and T Britton (2000) Stochastic Epidemic Models and their Statistical Analysis, Springer, New York.

\bibitem{B11} G Béraud, S Kazmercziak, P Beutels, D Levy-Bruhl, X Lenne et al. (2015). The French Connection: The First Large
Population-Based Contact Survey in France
Relevant for the Spread of Infectious
Diseases. \emph{PLoS One}, DOI:10.1371/journal.pone.0133203.

\bibitem{BBT} Britton, T., Ball F., Trapman P.\ (2020). A mathematical model reveals the influence of population heterogeneity on herd immunityto SARS-CoV2. \emph{Science}, 369 (6505), pp. 846-849.

\bibitem{CWG20} Coletti P, Wambua J, Gimma A, Willem L, et al. (2020). CoMix: comparing mixing patterns in the Belgian population during and after lockdown. \emph{Scientific Reports}, 10:21885

\bibitem{DHB13} O.~Diekmann, H.~Heesterbeek, T.~Britton, {\it Mathematical tools for  understanding infectious disease dynamics\/}, (Princeton University Press, 2013).

\bibitem{HCM19} Hoang, T., Coletti, P., Melegaro, A., Wallinga, J., Grijalva, C.G., Edmunds, J.\ et al. (2019). A systematic review of social contact surveys to inform transmission models of close-contact infections. \emph{Epidemiology}, 30: 723-736.

\bibitem{H11} Horby P, Thai PQ, Hens N, Yen NTT, Mai LQ, et al. (2011) Social Contact Patterns in Vietnam and Implications for the Control of Infectious Diseases. \emph{PLoS One} 6(2): e16965. doi:10.1371/journal.pone.0016965

\bibitem{L21} CY Liu, J Berlin, MC. Kiti, E Del Fava, A Grow et al. (2021). Rapid Review of Social Contact Patterns During the
COVID-19 Pandemic. \emph{Epidemiology} 32: 781–791)

\bibitem{M08} Mossong J, Hens N, Jit M, Beutels P, Auranen K, et al. (2008). Social Contacts and Mixing Patterns Relevant to the Spread of Infectious Diseases. \emph{PLOS Medicine} 5(3): e74.

\bibitem{SCD} http://www.socialcontactdata.org/data/


\bibitem{W12} Willem L, Van Kerckhove K, Chao DL, Hens N, Beutels P. (2012). A nice day for an infection? Weather conditions and social contact patterns relevant to influenza transmission. \emph{PloS One} 7(11):e48695.

\bibitem{WTK06} Wallinga J, Teunis and Kretzschmar. (2006). Using data on social contacts to estimate age-specific transmission parameters for respiratory-spread infectious agents. \emph{Am.\ J.\ Epidemiology}, 164: 936-944.

\bibitem{ZLH20} B Zhou, X Lu, P Holme. (2020). Universal evolution patterns of degree assortativity in social networks. \emph{Social Networks}, 63:47-55.

\end{thebibliography}
\end{document}